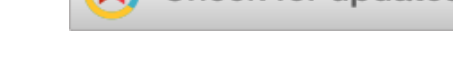

# A novel approach for identifying and addressing case-mix heterogeneity in individual participant data meta-analysis

Tat-Thang Vo[1,2] | Raphael Porcher[2] | Anna Chaimani[2] | Stijn Vansteelandt[1,3]

[1]Department of Applied Mathematics, Computer Science and Statistics, Ghent University, Ghent, Belgium

[2]Université de Paris, CRESS, INSERM, INRA, Paris, France

[3]Department of Medical Statistics, London School of Hygiene and Tropical Medicine, London, UK

**Correspondence**
Tat-Thang Vo, Department of Applied Mathematics, Computer Science and Statistics, Ghent University, Krijgslaan 281, S9, Ghent 9000, Belgium.
Email: tatthang.vo@ugent.be

Case-mix heterogeneity across studies complicates meta-analyses. As a result of this, treatments that are equally effective on patient subgroups may appear to have different effectiveness on patient populations with different case mix. It is therefore important that meta-analyses be explicit for what patient population they describe the treatment effect. To achieve this, we develop a new approach for meta-analysis of randomized clinical trials, which use individual patient data (IPD) from all trials to infer the treatment effect for the patient population in a given trial, based on direct standardization using either outcome regression (OCR) or inverse probability weighting (IPW). Accompanying random-effect meta-analysis models are developed. The new approach enables disentangling heterogeneity due to case mix from that due to beyond case-mix reasons.

## 1 | INTRODUCTION

Meta-analysis is a cornerstone of comparative effectiveness research, as it allows synthesizing the evidence from multiple randomized controlled trials (RCTs) and inferring the effect of interventions with increased precision.[1,2] A key issue in meta-analysis is heterogeneity, which arises due to the fact that studies included in a systematic review often differ to some degree in the case mix of participants, the variant of the intervention, settings, and outcome measurement.[3] In view of this, one of the first steps in every systematic review and meta-analysis is to define the target population of the meta-analysis as part of the population, intervention, control, and outcome (PICO) criteria for considering studies for inclusion.[2] Nevertheless, even when studies are chosen to be as similar as possible in terms of PICO, some amount of heterogeneity is usually inevitable, and it is often challenging to make decisions upon how much heterogeneity is acceptable.[2] Restricting the inclusion criteria not only would probably result in a sufficiently homogenous database but also limits the generalizability of the findings. In contrast, broadening the PICO criteria may result in increased heterogeneity.

In the presence of statistical heterogeneity, meta-analysts usually try to explain it using subgroup analyses and meta-regression. A common concern with these methods is the potential of aggregation bias. In particular, associations observed between the outcome and study characteristics across studies need not be present within studies, and vice versa.[2] This makes meta-analyses of individual participant data (IPD) preferable, when such data are available, as they allow to investigate the impact of different characteristics on the outcome both within and across studies.[4-6] There are two main approaches for IPD meta-analysis: (i) a two-stage approach where each study is first analyzed separately and the study-specific estimates are then combined using similar techniques as aggregate data meta-analysis; and (ii) a one-stage approach where all IPD from all studies are analyzed in a single model accounting for clustering.[7] Despite the advantage of IPD meta-analysis over meta-analysis of aggregate data to handle differences in the case mix of the studies, the classical approaches to IPD meta-analysis still have limitations.

Assume, for instance, that the intervention has a different effect in different patient subgroups (eg, patients with small and large lesions, as in the illustration presented in the Supporting Information S1) but the subgroup-specific effect is constant across studies. If the subgroups are differently distributed across the trials, then the intervention effect in each trial will be different. Upon pooling the different estimates, one is then likely to detect heterogeneity, despite the effect of the intervention being homogeneous across the studies. A one-stage or two-stage IPD meta-analysis can adjust for the prognostic value of the lesion size but then returns subgroup effects.[8-10] In certain situations, subgroup effects might be of less interest than the population treatment effect. In particular, in the analysis of binary or time-to-event outcomes, effect measures like odds ratios and hazard ratios are well known to suggest larger intervention effects when being calculated for subgroups, compared with when being calculated for the population (even when all subgroups experience the same effect on the odds or hazard ratio scale).[11]

In this paper, we propose a new approach for IPD meta-analysis of RCTs that allows (a) to control for differences in the case mix across studies and reduce heterogeneity and (b) to infer the treatment effect for a population that is well defined in terms of case mix. Building on recent work by Bareinboim and Pearl,[12,13] this is achieved by standardizing the results from the different trials to the same patient population, eg, the patient population observed in one of the trials or any other population of interest, before meta-analyzing them as in a classical two-stage IPD meta-analysis.[5,6] As an added advantage, this enables one to decompose the overall heterogeneity between the trial results into two different sources, which the usual approaches to IPD meta-analysis do not provide: "case-mix heterogeneity" (ie, arising when the treatment effect is modified by one or more of the factors used to define case mix) and "beyond case-mix heterogeneity" (ie, arising due to the difference between studies in design or methodological aspects).

We proceed as follows. In Section 2, we propose two estimators that aim to standardize results of different trials over the case mix of a target population. The subsequent meta-analysis then infers the treatment effect in the given population by using the outcome data standardized from other trials. We show in Section 3 that under certain conditions, this approach not only gives valid results but also allows for a more insightful assessment of heterogeneity in meta-analysis. The novel approach is illustrated by reanalyzing a published IPD meta-analysis evaluating the effect of vitamin D on the risk of respiratory infection in Section 4. Some important challenges are then extensively discussed in Section 5.

**WHAT IS ALREADY KNOWN**

- Meta-analysis of individual participant data (IPD-MA) is considered to be a gold standard of systematic review.
- A one-stage or two-stage IPD-MA can adjust for the prognostic and predictive value of different baseline characteristics but often returns subgroup effects.

**WHAT IS NEW**

- We propose a new approach for IPD-MA of randomized controlled trials (RCTs)that allows one (a) to control for differences in the case mix across studies and reduce heterogeneity and (b) to infer the treatment effect for a population that is well defined in terms of case mix.
- The overall heterogeneity across trials is decomposed into case-mix heterogeneity and beyond case-mix heterogeneity.

**POTENTIAL IMPACT FOR RSM READERS**

- In the original trial reports, trialists may consider producing an effect measure estimate standardized to the case-mix distribution of a reference population. Meta-analysts could then base a standard meta-analysis on these mutually standardized estimates, which would have the advantage of describing the effect for the same population. This would overcome the need for an IPD-MA.

## 2 | CASE-MIX STANDARDIZATION IN META-ANALYSIS OF RANDOMIZED CONTROLLED TRIALS: A CAUSAL FRAMEWORK

### 2.1 | Setting

Consider a meta-analysis of $K$ RCTs to evaluate the comparative effectiveness of two treatments ($X = 1$ vs 0) on a dichotomous outcome $Y$ (1 vs 0). Let $S$ be an indicator of the study from which a given patient originates, which takes values from 1 to $K$. To make the beyond case-mix difference between studies explicit, we will label the versions of treatment $x$ as $x_1$ to $x_K$ (ie,$x = 0,1$) for studies 1 to $K$, respectively. Note that even when the same treatment is evaluated across studies, the version of treatment will still likely be different, eg, because of differences in standard of care or patient management between studies or because in one study, there is a greater attempt to prevent noncompliance than in other studies. Besides, we denote $Y(x_k)$ as the outcome that would be observed in a patient if this patient were assigned to the version of treatment $x_k$. Each patient, therefore, will have $2K$ potentially observed outcomes. However, since each patient is only assigned to one specific version of treatment or control, not all of these outcomes can actually be observed for each patient in practice. Due to this, the proposed outcomes $Y(x_k)$ are often referred to as counterfactual outcomes. A more detailed discussion about the counterfactual outcome framework can be found elsewhere.[14,15]

Let $P\{Y(x_k) = 1 | S = j\}$ $(x = 0{,}1)$ denote the chance of success if the patients in population $j$ were given the version of treatment/control used in study $k$. On the basis of these probabilities, the effect of the treatment version $k$ in population $j$ can be expressed as a risk difference, relative risk, or odds ratio. For instance, on the relative risk scale, we denote the following:

$$RR(j,k) = \frac{P\{Y(1_k) = 1 | S = j\}}{P\{Y(0_k) = 1 | S = j\}},$$

which expresses the treatment effect when all individuals from population $S = j$ were given the (version of) treatment versus control used in trial $k$. As discussed below, the effects $RR(j,k)$ for different $k = 1,\ldots,K$, but the same $j$ are potentially more homogeneous, since the case-mix heterogeneity is canceled out and all $RR(j,k)$ describe the treatment effects for the same population $j$.

## 2.2 | Assumptions

To identify $RR(j,k)$ and the corresponding probabilities, the following assumptions are made:

a Ignorable study assignment,[12,13] which states that the trial indicator is independent of all counterfactual outcomes, conditioning on the prognostic factors $L$; that is, $Y(x_k) \perp S \mid L$ for $x = 0{,}1$ and $k = 1,\ldots,K$, where $A \perp B \mid C$ for random variables $A$, $B$, and $C$ means that $A$ is conditionally independent of $B$, given $C$. This implies that individuals with the same characteristics $L$ in different trials would have the same outcome risks if given the same treatments. This is satisfied when $L$ contains all prognostic factors of the outcome that are differentially distributed between studies. This assumption cannot be tested in practice. However, it is partially testable when the control is the same in different studies, in the sense that $Y(0_1) = Y(0_2) = \ldots = Y(0_K) = Y(0)$, for then, it should imply that $Y \perp S \mid X = 0, L$, which is testable. In practice, when there is evidence against the assumption that outcome is independent of trial indicator given $X = 0$ and $L$, one should first carefully verify the added assumption of common control (eg, whether the control groups in different trials are really similar in terms of pharmacological properties or of associated risks of bias). If this is indeed the case, then the considered set of covariates $L$ is likely insufficient to define the case mix of the included studies. Such a limitation should be acknowledged. Note that standard meta-analysis approaches are also biased when this assumption is violated. This is because summaries over studies that include very different case mix are prone to bias, as explained in Section 1, unless they involve an appropriate case-mix adjustment.

b Positivity,[16] which states that any individual with characteristics $L_i$ in study $S_i = k$ has a positive probability, based on these characteristics, of being included in study $j$: $0 < P(S_i = j | L_i) < 1$. Violations of positivity may be deterministic or random.[16] A deterministic violation occurs when the target populations of trials are relatively different; eg, one study only includes children, whereas the others recruit adults. In contrast, random violations of positivity may occur when there are trials of small sample size. In that case, it may happen by chance that no one in a given age class participates in one trial, even though the restrictions on age for eligibility are the same across trials. Besides, note that what is meant by positivity here is different from the conventional positivity assumption that appears in causal theory.[16] The former basically assumes that $P(S = j | L)$ for patients in trial $k$ is nonzero, which guarantees an adequate overlap between different trial populations in terms of case mix. This is important to be able to learn about the treatment effect in the target population from what is observed in the original one.

c Consistency,[17] which states that $Y(x_k)$ agrees with the observed outcome $Y$ for all individuals in study $k$ ($k=1,\ldots,K$) who received treatment $x$ ($x = 0{,}1$). This assumption is generally plausible in randomized trials.[18]

d Ignorable treatment assignment within study,[17] which states that within each trial, the treatment is independent of all counterfactual outcomes $-Y(x_k) \perp X \mid S$ for $x = 0{,}1$. This assumption is guaranteed to hold because of randomization within each individual trial.[19]

## 2.3 | Outcome regression approach

Under the aforementioned assumptions, it can be shown (Supporting Information S2) that

$$P\{Y(x_k) = 1 | S = j\} = \mathrm{E}[\mathrm{E}(Y | X = x, L, S = k) | S = j]$$

$$= \sum_{l} P(Y=1|X=x, L=l, S=k) \times P(L=l|S=j).$$

Intuitively, this formula amounts to a simple recalibration (or reweighting) of the $L$-specific effects to account for the new $L$'s distribution.[12] Assume that in population $k$, the outcome for each patient follows a logistic model:

$$P(Y=1|X,L,S=k) = \text{expit}(\beta_{0k} + \beta_{1k}X + \beta_{2k}L + \beta_{3k}XL), \quad (1)$$

where expit $(a)$ = $\{1+exp(-a)\}^{-1}$. Under Model (1), a straightforward estimator of $P\{Y(x_k) = 1|\, S = j\}$ is obtained by using outcome regression (OCR):

$$\hat{P}\{Y(x_k)=1|S=j\} = \frac{\sum_i I(S_i=j)\text{expit}\left(\hat{\beta}_{0k} + \hat{\beta}_{1k}x + \hat{\beta}_{2k}L_i + \hat{\beta}_{3k}xL_i\right)}{\sum_i I(S_i=j)}.$$

As a result, $RR(j,k)$ can be estimated as follows:

$$\hat{RR}(j,k) = \frac{\sum_i I(S_i=j)\text{expit}\left(\hat{\beta}_{0k} + \hat{\beta}_{1k} + \hat{\beta}_{2k}L_i + \hat{\beta}_{3k}L_i\right)}{\sum_i I(S_i=j)\text{expit}\left(\hat{\beta}_{0k} + \hat{\beta}_{2k}L_i\right)},$$

where $\hat{\beta}_{0k}, \hat{\beta}_{1k}, \hat{\beta}_{2k}$, and $\hat{\beta}_{3k}$ are estimates obtained by fitting Model (1) to the data from trial $k$. This strategy suffers from two drawbacks. First, the result of transporting the findings across studies may be heavily dependent upon the choice of model for the outcome, eg, on the decision to include interactions of treatment with some baseline covariates. Second, this approach comes with a high risk for extrapolation when patients in different studies have very different case mix.[20] Such extrapolation is the result of making the outcome model fit well over the case mix of study $k$ but then using it to make outcome predictions for the possibly different case mix in study $j$. The severity of that extrapolation may easily go unnoticed in practice.

## 2.4 | Inverse probability weighting approach

In view of the aforementioned concerns, we considered an alternative approach based on inverse probability weighting (IPW). IPW is a method commonly used to obtain marginal effects in observational studies, especially when there is time-dependent confounding.[21,22] It can be shown (Supporting Information S2) that

$$P\{Y(x_k)=1|S=j\} = \frac{1}{P(S=j)} \mathrm{E} \left\{ I(S=k).Y.I(X=x).\frac{P(S=j|L)}{P(S=k|L)}.\frac{1}{P(X=x|S=k)} \right\}.$$

Assume further that for a given patient with the covariate profile $L$, the probability to be in trial $j$ vs in trial $k$ follows a logistic propensity score (PS) model:

$$\frac{P(S=j|L)}{P(S=k|L)} = \text{expit}(\gamma_0 + \gamma_1 L). \quad (2)$$

This suggests estimating $P\{Y(x_k) = 1|\, S = j\}$ as follows:

$$\hat{P}\{Y(x_k)=1|S=j\} = \frac{\sum_i \left\{ I(S_i=k).Y_i.I(X_i=x).\text{expit}(\hat{\gamma}_0 + \hat{\gamma}_1 L_i).\frac{1}{\hat{P}(X_i=x|S_i=k)} \right\}}{\sum_i I(S_i=j)},$$

where $\hat{\gamma}_0$ and $\hat{\gamma}_1$ are the estimates obtained by fitting Model (2) to the data from trials $j$ and $k$. This results in the following estimator for $RR(j,k)$:

$$\hat{RR}(j,k) = \frac{1}{R_k} \times \frac{\sum_i I(S_i=k) Y_i X_i \text{expit}(\hat{\gamma}_0 + \hat{\gamma}_1 L_i)}{\sum_i I(S_i=k) Y_i (1-X_i) \text{expit}(\hat{\gamma}_0 + \hat{\gamma}_1 L_i)},$$

where $R_k$ is the ratio between the number of treated vs untreated patients in the trial $k$. Calculating this requires no modeling assumption about the outcome generating mechanism. Therefore, the estimator does not require a model for the outcome, which is important because huge extrapolations could otherwise be made if the outcome model ignored certain forms of heterogeneity (eg, covariate by study interactions). Instead, a PS model for $P(S = j|\, L)$ now must be correctly specified (eg, by using multinomial regressions) to ensure that the estimator is unbiased in sufficiently large samples.[21-23]

The IPW approach can be susceptible to the presence of unstable weights, that is, to some weights $\frac{\hat{P}(S_i=j|L_i)}{\hat{P}(S_i=k|L_i)} = \text{expit}(\hat{\gamma}_0 + \hat{\gamma}_1 L_i)$ being very large for some individuals.[22] The estimation by IPW is then dominated by these large weights, which results in a huge reduction in effective sample size.[22,24] In extreme cases, the IPW estimate for $P\{Y(x_k) = 1|\, S = j\}$ can even exceed the theoretical boundary of 1. This can be remedied by noting that (see Supporting Information S2)

$$P\{Y(x_k)=1|S=j\} = \frac{\mathrm{E}\left\{ I(S=k).Y.I(X=x).\frac{P(S=j|L)}{P(S=k|L)} \right\}}{\mathrm{E}\left\{ I(S=k).I(X=x).\frac{P(S=j|L)}{P(S=k|L)} \right\}},$$

which suggests alternatively estimating $RR(j,k)$ as follows:

$$\hat{RR}(j,k) = \frac{\frac{\sum_i I(S_i = k) Y_i X_i \text{expit}(\hat{\gamma}_0 + \hat{\gamma}_1 L_i)}{\sum_i I(S_i = k) X_i \text{expit}(\hat{\gamma}_0 + \hat{\gamma}_1 L_i)}}{\frac{\sum_i I(S_i = k) Y_i (1 - X_i) \text{expit}(\hat{\gamma}_0 + \hat{\gamma}_1 L_i)}{\sum_i I(S_i = k)(1 - X_i) \text{expit}(\hat{\gamma}_0 + \hat{\gamma}_1 L_i)}}.$$

The additional denominators ensure weight stabilization, in the sense that they deliver weights between 0 and 1, thereby preventing the resulting stabilized IPW estimate for $P\{Y(x_k) = 1 | S = j\}$ from exceeding the boundary[22] of 1. Extreme weights will often appear in settings where the different trials consider very different case mix. They thus give the user a warning that it can be tricky to pool the results from such different trials, which could go unnoticed with the OCR approach as well as with the standard meta-analysis approach.

Other effect measures (such as risk difference and odds ratio) can also be defined and estimated in a similar way. The definition of the odds ratio $OR(j,k)$ and its two corresponding estimators is given in Supporting Information S3.

## 2.5 | Deriving summary estimates and dismantling the two sources of heterogeneity

To summarize the results $\hat{RR}(j,k)$ obtained from the same population $j$, a random effect meta-analysis of the form

$$\log\left(\hat{RR}(j,k)\right) \sim \mathcal{N}\left(\log\left(RR(j,k)\right) \sigma_{jk}^2\right),$$

$$\log\left(RR(j,k)\right) \sim \mathcal{N}\left(\log\left(RR(j.)\right), \vartheta_j^2\right),$$

may now be performed. Here, $\log(RR(j.))$ expresses the treatment effect for the target population $j$, which can then be estimated via a weighted average of the log relative risks $\log\left(\hat{RR}(j,k)\right)$:

$$\log\left(\hat{RR}(j.)\right) = \frac{\sum_{j=1}^{K} w_j . \log\left(\hat{RR}(j,k)\right)}{\sum_{j=1}^{K} w_j} \text{ with } w_j = \frac{1}{\hat{\sigma}_{jk}^2 + \hat{\vartheta}_j^2}.$$

This pooled estimate describes the summary treatment effect for the underlying population $j$. The standard error $\sigma_{jk}$ of $\log\left(\hat{RR}(j,k)\right)$ can be estimated by either bootstrap or sandwich estimators. Further, $\vartheta_j^2$ expresses how much results from different trials vary even when considered for the same patient population. This may result, for instance, from the differential effect of different treatment versions in the different trials. Importantly, since all estimates $\hat{RR}(j,k)$ focus on the same patient population (in terms of covariates $L$), $\vartheta_j^2$ does not express heterogeneity due to differential case mix (in covariates $L$).

An added advantage of the proposed framework is that, in view of the above, it results in a more informative way of assessing heterogeneity. Indeed, since different $RR(j,k)$ of the same population $j$ are standardized over the same covariate distribution, these may only be heterogeneous due to beyond case-mix reasons across the different population $k$. As a result, by testing the equality of $RR(j,k)$, $k = 1,\ldots,K$, for the same population $j$, one may develop insight in beyond case-mix heterogeneity. Similarly, when different $RR\ (j,k)$, $j = 1,\ldots,K$, differ for the same population $k$, there is heterogeneity due to differential case mix among the populations.

Comparison among different $RR(j,k)$ can be done by using a Wald test. Consider, for instance, the null hypothesis $H_0 : RR(j,1) = RR(j,2) = \ldots = RR(j,K)$, which states that there is no beyond case-mix heterogeneity. This can be rewritten in matrix form as $H_0 : \boldsymbol{M}.\ \boldsymbol{RR} = \boldsymbol{0}$, where

$$\boldsymbol{RR} = \left(\boldsymbol{RR}(\boldsymbol{1}), \boldsymbol{RR}(\boldsymbol{2}), \ldots, \boldsymbol{RR}(\boldsymbol{K})\right)^t,$$

$\boldsymbol{RR}(\boldsymbol{j}) = (\log RR(j,1), \log RR(j,2), \ldots, \log RR(j,K))$,

and $\boldsymbol{M}$ is an appropriately chosen $(K - 1) \times K^2$ matrix. Under this null hypothesis, the test statistic

$$\boldsymbol{T} = \left(\boldsymbol{M}.\hat{\boldsymbol{RR}}\right)^t \left[\mathbf{M}\left(\hat{\sum}\right)\mathbf{M}^t\right]^{-1} \left(\boldsymbol{M}.\hat{\boldsymbol{RR}}\right) \sim \chi_{K-1}^2,$$

where $\hat{\sum}$ is the estimate of the covariance matrix of $\hat{\boldsymbol{RR}}$. Here, $\hat{\sum}$ is derived by using conventional methods like bootstrap or sandwich estimators.[25]

As a final remark, note that when all trials have the same control treatment, the assumption $Y \perp S \mid X = 0,L$ naturally implies that beyond case-mix heterogeneity can be interpreted as treatment effectiveness heterogeneity. Indeed, supposing, for instance, that the outcome generating mechanism in population $k$ ($k = 1, \ldots, K$) obeys the logistic model (1), for then,

$$P(Y = 1 | X = 0, L, S = k) = \text{expit}(\beta_{0k} + \beta_{2k} L).$$

The assumption that $Y \perp S \mid X = 0,L$ then implies that $\text{expit}(\beta_{01}+\beta_{21}L) = \ldots = \text{expit}(\beta_{0K}+\beta_{2K}L)$ for each $L$. This holds if and only if $\beta_{01} = \ldots = \beta_{0K}$ and $\beta_{21} = \ldots = \beta_{2K}$, which means that all coefficients that are not related to the treatment must be constant over different studies. Beyond case-mix heterogeneity, if present, is then due to differential treatment-related coefficients across populations.

## 3 | A SIMULATION STUDY

### 3.1 | Design

We apply the proposed methods in numerically simulated meta-analyses of RCTs that evaluate a (binary) active treatment ($X$) versus control with respect to a binary outcome ($Y$). We consider five settings. In each setting, the meta-analysis includes five trials with a total of 3750 patients. In settings 1 to 4, the impact of treatment $X$ and of five continuous outcome predictors $L_i$ ($i = 1, \ldots, 5$) on $Y$ in population $k$ ($k = 1, \ldots, 5$) is generated using the following logistic models:

$$P(Y=1|X,\mathbf{L},S=k) = \text{expit}\left(\beta_0 + \beta_{1k}X + \sum_{i=1}^{5}\beta_{2i}L_i + \beta_{3k}X.L_1\right),$$

in line with the discussion in the previous section. The control group is assumed to be similar in the five trials, in the sense that $Y \perp S \mid X = 0,L$. This implies that $\beta_0$ and $\beta_{2i}$ ($i = 1, \ldots, 5$) are fixed across the trials. In contrast, the two coefficients $\beta_{1k}$ and $\beta_{3k}$ take different values in each different population $k$.

The trial indicator $S$ is generated using a multinomial logistic model in the first three settings (see Table 1 for details). In setting 4, all five trials share a similar target

**TABLE 1** The mathematical symbols in this table are not well written. Please see attached how we want this table to look like. AllNumerical setup of the simulation study

| Setting | Numerical Setup |
|---|---|
| 1 | $\ell = (-0.5\quad -0.25\quad 0.00\quad 0.25\quad 0.50)^t$ $\quad\mathbf{\Sigma} = \text{diag}(0.50,\ 0. 50,\ 0.50,\ 0.50,\ 0.50)$<br>$\boldsymbol{M} = (1\quad L_1\quad L_2\quad L_3\quad L_4\quad L_5\quad L_1 \times L_2)$ $\quad\boldsymbol{\beta} = \begin{pmatrix} 0.01 & 0.45 & 0.30 & 0.30 & 0.30 & 0.01 & 0.01 \\ 0.01 & -0.20 & 0.20 & 0.20 & 0.20 & 0.50 & -2.25 \\ 0.01 & 0.45 & 0.10 & 0.10 & 0.10 & 0.50 & -2.25 \\ -0.25 & -0.55 & -0.25 & 0.55 & -0.45 & 0.25 & -1.50 \end{pmatrix}$<br>$\boldsymbol{P} = (1\quad X\quad L_1\quad L_2\quad L_3\quad L_4\quad L_5\quad X \times L_1)$<br>$\boldsymbol{\gamma} = (-0.25\quad -0.5\quad 0.5\quad -0.25\quad -0.5\quad 0.25\quad -0.5\quad -1.5)$<br>$RR(1, 1) = 1.30$; $RR(2, 2) = 1.08$; $RR(3, 3) = 1.20$; $RR(4, 4) = 0.91$; $RR(5, 5) = 1.32$ |
| 2 | $\ell = (-0.5\quad -0.25\quad 0.00\quad 0.25\quad 0.50)^t$ $\quad\mathbf{\Sigma} = \text{diag}(0.50,\ 0. 50,\ 0.50,\ 0.50,\ 0.50)$<br>$\boldsymbol{M} = (1\quad L_1\quad L_2\quad L_3\quad L_4\quad L_5\quad L_1 \times L_2)$ $\quad\boldsymbol{\beta} = \begin{pmatrix} 0.25 & 1.00 & 0.55 & -0.25 & 0.75 & -0.25 & -0.5 \\ 0.25 & 0.35 & 0.45 & -0.35 & 0.65 & 0.25 & 0.5 \\ 0.25 & 1.00 & 0.35 & -0.45 & 0.55 & 0.25 & 0.5 \\ 0.25 & 0.55 & 0.25 & -0.55 & 0.45 & -0.25 & -0.5 \end{pmatrix}$<br>$\boldsymbol{P} = (1\quad X\quad L_1\quad L_2\quad L_3\quad L_4\quad L_5\quad X \times L_1)$<br>$\boldsymbol{\gamma} = (-0.25\quad \beta_{1k}\quad 0.5\quad -0.25\quad -0.5\quad 0.25\quad -0.5\quad -1.5)$<br>$\beta_{1k} = -0.25 \times I(S = 1) + 0.65 \times I(S = 2) + 0 \times I(S = 3) + 0.5 \times I(S = 4) + 0.25 \times I(S = 5)$<br>$RR(1, 1) = 1.55$; $RR(2, 2) = 1.58$; $RR(3, 3) = 1.60$; $RR(4, 4) = 1.57$; $RR(5, 5) = 1.52$ |
| 3 | $\ell = (0.5\quad -0.25\quad 0.00\quad 0.25\quad 0.50)^t$ $\quad\mathbf{\Sigma} = \text{diag}(7.50,\ 2. 00,\ 2.00,\ 2.00,\ 2.00)$<br>$\boldsymbol{M} = (1\quad L_1\quad L_2\quad L_3\quad L_4\quad L_5\quad L_1 \times L_1)$ $\quad\boldsymbol{\beta} = \begin{pmatrix} -0.14 & -1.16 & 0.20 & -0.20 & 0.20 & -0.20 & 1.18 \\ -1.60 & -2.02 & 0.20 & -0.40 & 0.40 & 0.30 & 2.04 \\ -1.46 & -2.00 & 0.30 & -0.30 & 0.30 & 0.20 & 2.03 \\ -1.33 & -2.00 & 0.20 & -0.20 & 0.40 & -0.30 & 2.02 \end{pmatrix}$<br>$\boldsymbol{P} = (1\quad X\quad L_1\quad L_2\quad L_3\quad L_4\quad L_5\quad X \times L_1)$<br>$\boldsymbol{\gamma} = (-0.25\quad -0.5\quad 0.5\quad -0.25\quad -0.5\quad 0.25\quad -0.5\quad -2)$<br>$RR(1, 1) = 0.48$; $RR(2, 2) = 0.55$; $RR(3, 3) = 0.86$; $RR(4, 4) = 0.82$; $RR(5, 5) = 0.81$ |
| 4 | $\ell = (-0.5\quad -0.25\quad 0.00\quad 0.25\quad 0.50)^t$ $\quad\mathbf{\Sigma} = \text{diag}(0.50,\ 0. 50,\ 0.50,\ 0.50,\ 0.50)$<br>$\boldsymbol{M} = \mathbf{0}$ $\quad\boldsymbol{\beta} = 0$<br>$\boldsymbol{P} = (1\quad X\quad L_1\quad L_2\quad L_3\quad L_4\quad L_5\quad X \times L_1)$<br>$\boldsymbol{\gamma} = (-0.25\quad \beta_{1k}\quad 0.5\quad -0.25\quad -0.5\quad 0.25\quad -0.5\quad -1.5)$<br>$\beta_{1k} = 0.25 \times I(S = 1) + 0.05 \times I(S = 2) - 0.15 \times I(S = 3) - 0.35 \times I(S = 4) - 0.50 \times I(S = 5)$<br>$RR(1, 1) = 1.60$; $RR(2, 2) = 1.49$; $RR(3, 3) = 1.37$; $RR(4, 4) = 1.25$; $RR(5, 5) = 1.16$ |
| 5 | $L_1 \mid S = 1 \sim Unif(0,0.50)$; $L_1 \mid S = 2 \sim Unif(-1.5,0.20)$; $L_1 \mid S = 3 \sim Unif(-0.20,0.45)$;<br>$L_1 \mid S = 4 \sim Unif(-0.15,0.40)$; $L_1 \mid S = 5 \sim Unif(-0.20,0.50)$;<br>$\boldsymbol{P} = (1\quad X\quad L_1\quad L_1^2\quad L_1^3\quad X \times L_1)$<br>$\boldsymbol{\gamma} = (1\quad -0.75\quad 1\quad 2\quad 2\quad 1)$<br>$RR(1, 1) = 0.89$; $RR(2, 2) = 0.55$; $RR(3, 3) = 0.82$; $RR(4, 4) = 0.79$; $RR(5, 5) = 0.84$ |

*Note*. From settings 1 to 4, we first generate the covariate vector $\boldsymbol{L} = (L_1 L_2 L_3 L_4 L_5)^t$ by using the multivariate normal distribution $N(\boldsymbol{\ell}, \sum)$. The trial indicator $S$ is then generated by using the multinomial model $\log\left\{\frac{P(S=j|\boldsymbol{L})}{P(S=1|\boldsymbol{L})}\right\} = \beta_{.j}\boldsymbol{M}^t$ for $j = 2, \ldots, 5$, where $\boldsymbol{M}$ is specific for each setting and $\boldsymbol{\beta}_{.j}$ is the $j$th row of $\boldsymbol{\beta}$. In setting 5, the covariate $L_1$ in each study is generated by a separate uniform distribution. Across settings, the outcome $Y$ in each trial is then generated by using a logistic model, ie, $\text{logit}P(Y = 1 \mid X,\boldsymbol{L},S) = lp$, where $lp = \boldsymbol{\gamma}\boldsymbol{P}^t$ with $\boldsymbol{\gamma}$ and $\boldsymbol{P}$ specific for each setting population and hence have the same case mix. In settings 1 and 3, the treatment effect is heterogeneous on the population level, although it is actually equally beneficial for patients with the same covariate profile regardless of their origin. While all assumptions (see Section 2.2) are fulfilled in setting 1, setting 3 assesses the behavior of the two estimators when the positivity assumption is nearly violated (ie, in Figure 1A, individuals with extreme values of $L_1$ are nearly never recruited in the first two trials).

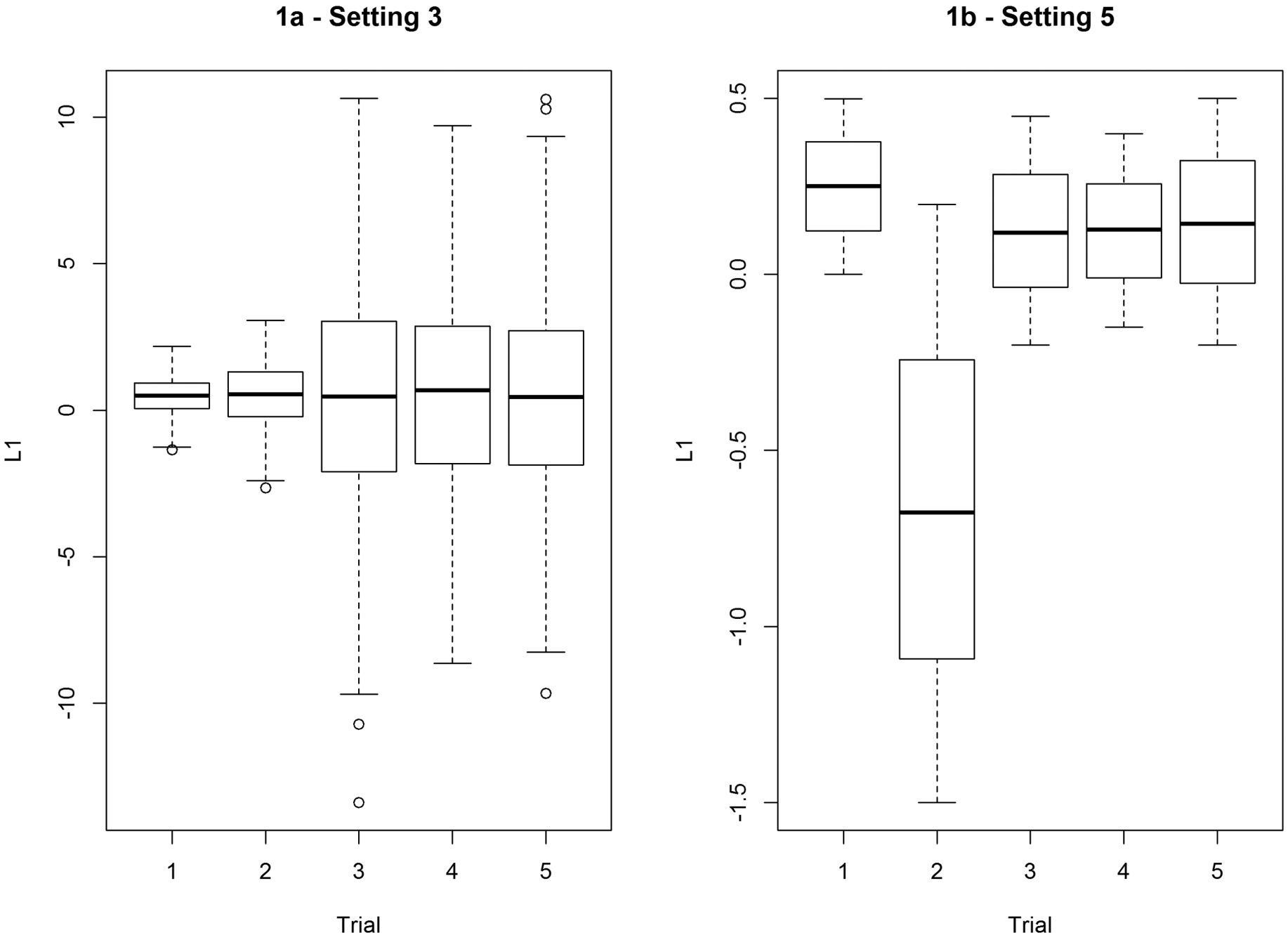

**FIGURE 1** Simulation study: the distribution of $L_1$ across the five trials in settings 3 (A) and 5 (B)

In setting 2, both case-mix heterogeneity and beyond case-mix heterogeneity are present, but the two sources of heterogeneity compensate each other and result in 5 (approximately) similar $RR(j,j)$ ($j = 1, \dots, 5$) across studies (Table 1). In setting 4, the five trials have nontrivial beyond case-mix heterogeneity.

We dedicate the final setting 5 to illustrate the risk of extrapolation when using the OCR approach. In this setting, we let one baseline characteristic $L_1$ be differently distributed among studies. Moreover, the case mix of the trial $S = 2$ is chosen to be considerably different from that of the remaining four trials (see Table 1 and Figure 1B). The impact of treatment and of the baseline factor $L_1$ (as well as $L_1^2$ and $L_1^3$) on the outcome is generated using a logistic model that is identical across the five trials (Table 1).

## 3.2 | Analysis

### 3.2.1 | Bias

The new meta-analysis approach is assessed on both probability and relative risk scales. Note that although the outcome obeys a logistic model, the use of relative risks is valid since we merely evaluate population relative risks. Across the first four settings, we investigate the bias of the two estimators when the logistic outcome model (for OCR-based approach) and the multinomial PS model (for IPW-based approach) are correctly specified. In setting 1, the two estimators are further assessed when the PS model (for the IPW-based estimators) is misspecified by not including the interaction between $L_1$ and $L_2$ and the outcome models (for the OCR-based estimators) are misspecified by not including the interaction between $X$ and $L_1$ (Table 1).

In setting 5, a logistic model without the cubic term $L_1^3$ is used for the OCR-based estimator that transfers the information from population 1 toward the other populations. In contrast, the outcome model is correctly specified when using the OCR-based estimator to transfer results of other trials toward trial 1, and the PS model only includes the main term of $L_1$ when using the IPW-based estimator (Table 1).

In each setting, the true values ($\theta$) of all estimands, namely, (a) $P\{Y(x_k) = 1 | S = j\}$ with $x = 0,1$ and (b) $RR(j,k)$, are derived as the average result across an independent 5000-run simulation, which make use of the true model coefficients. For instance, in settings 1 to 4, the true value for $P\{Y(x_k) = 1 | S = j\}$ with $x = 0,1$ is calculated as $\frac{1}{5000}\sum_{m=1}^{5000}\hat{P}_m$, where

$$\hat{P}_m = \frac{\sum_{i=1}^{3750} I(S_i = j)\,\text{expit}\left(\beta_0 + \beta_{1k}x + \sum_{a=1}^{5}\beta_{2a}L_{a,i} + \beta_{3k}x.L_1\right)}{\sum_{i=1}^{3750} I(S_i = j)}.$$

The mean ($\bar{\theta}$) of the corresponding OCR/IPW-based estimator is computed over the main simulations. The bias and the relative bias are computed as $(\bar{\theta}-\theta)$ and $(\bar{\theta}-\theta)/\theta$, respectively.

### 3.2.2 | Summary estimates and heterogeneity assessment

The variance of the two proposed estimators for $\log RR(j,k)$ is derived using sandwich estimators.[25] After case-mix standardization, the results $\log \hat{RR}(j,k)$ obtained from the same population $j$ are summarized by a random-effect model specific for population $j$ (see Section 2.5). The between-trial variances, the $I^2$ statistics, and the performance of the heterogeneity tests conducted after case-mix standardization are compared with when a standard two-step meta-analysis of $\log \hat{RR}(j,j)$ $(j=1,\ldots,5)$ without covariate adjustment is conducted. The comparison of different relative risks is realized using Wald tests.

## 3.3 | Result

### 3.3.1 | Bias

The evaluation of bias when estimating $P\{Y(1_k)=1 | S=j\}$ and $RR(j,k)$ is presented in Table 2. In setting 1, the OCR-based estimator yields no bias when the outcome model is correctly specified. In contrast, the IPW-based estimator is slightly biased when standardizing results of trials 3 and 4 over the case mix of trial 1, although the PS model is correctly specified. This can be explained by the fact that patients with large absolute values of both $L_1$ and $L_2$ are more likely to be recruited in trial 3 (and 4) than in trial 1 (Table 1). These patients are then strongly weighted and influence the IPW-based estimator. While this minor violation of positivity is flagged by the presence of the large weights, it is overlooked by the OCR-based estimator. Besides, both estimators are biased when the essential interactions are not included in the PS model or the outcome model (setting 1, Table 2).

In setting 2, both estimators behave properly. In setting 3, the information from trial 1 cannot be standardized over the case mix of trials 3 to 5 via an IPW-based estimator (although the PS model is correctly specified). In fact, the IPW-based estimator is strongly driven by outcomes occurring in some patients with extremely large weights. This is most clearly seen in estimates of the probabilities $P\{Y(x_k)=1 | S=j\}$, which in some simulations exceed the boundary of 1. Roughly speaking, to standardize the results of one study over the case mix of the other, we learn from subjects in different studies with similar characteristics. As individuals with extreme values of $L_1$ present only in trials 3 to 5, there is no information about the effect of treatment assignment in trial 1 (and 2) for these individuals. Such lack of information becomes apparent through the unstable behavior of the IPW-based estimator (Table 2). While this is not problematic for the OCR-based estimator when the model is correctly specified, it does rely on extreme extrapolation.

In the same setting 3, the IPW-based estimator is slightly biased when standardizing the results of trials 3 to 5 over the case mix of trial 1 (Table 2). The reason is that patients with $L_1$ value being closed to 0 will have a higher chance to be recruited in trials 3 to 5 than in trial 1 (see Table 1). The weights of these patients, therefore, can be fairly large. In contrast, such a fairly weak overlap between the two trials' case mix is not notified by the OCR-based estimator.

In setting 4, the two estimators both give valid results when the involved models are correctly specified. In setting 5, the OCR-based estimator standardizing the results of trial 1 over the case mix of population 2 is biased (Table 2). In fact, the model without the cubic term $L_1^3$ properly fits the data in trial 1. However, as there is little overlap between the two populations, using such a model for prediction in study 2 results in severe extrapolation. The OCR approach simply ignores such concern and hence yields relatively severe bias. In contrast, as the outcome model is correctly specified when transporting the information from trial 2 (which also has a much more heterogeneous case mix) toward population 1, the OCR-based estimator has no bias. While using the IPW approach also results in bias as the PS model is incorrectly specified, the presence of extreme weights could at least provide an alert on such bias and on the nonoverlap between trials in terms of case mix.

### 3.3.2 | Summary estimates and heterogeneity assessment

The summary estimates derived from the population-specific meta-analyses are provided in Table 3. These summaries have a larger variance when the case-mix standardization is conducted by using the IPW approach.

As can be seen from Tables 3 and 4, the proposed approaches correctly specify the source(s) of the total heterogeneity when the two estimators behave properly. For instance, the population-specific meta-analyses in setting 1 return a between-trial variance and $I^2$ statistic of 0, which indicates correctly that no heterogeneity presents after case-mix standardization (Table 4). The OCR-based tests assessing the beyond case-mix heterogeneity in setting 1 also show statistical significance in only 5% of the simulations, which is the conventional level of type I error. Across the settings, the proposed tests are more powerful when using the OCR-based estimator. In setting 2, simply meta-analyzing $\log RR(j,j)$ may suggest that no heterogeneity is present. In settings 1, 3, and 5, a standard heterogeneity assessment correctly detects the

**TABLE 2** Simulation results: bias assessment for $P\{Y(1_k) = 1 | S = j\}$ and $RR(j,k)$

| | $j = 1$ | | | | $k = 1$ | | | |
|---|---|---|---|---|---|---|---|---|
| **Setting**[a] | $k = 2$ | $k = 3$ | $k = 4$ | $k = 5$ | $j = 2$ | $j = 3$ | $j = 4$ | $j = 5$ |
| 1(P), OCR | 0.0005 (0.1) | 0.0000 (0.0) | 0.0004 (0.1) | 0.0000 (0.0) | 0.0003 (0.1) | −0.0006 (−0.2) | −0.0007 (−0.2) | 0.0000 (0.0) |
| 1(P), IPW | 0.0006 (0.1) | 0.0028 (0.6) | −0.0059 (−1.2) | −0.0011 (−0.2) | 0.0002 (0.1) | −0.0008 (−0.2) | −0.0010 (−0.3) | 0.0004 (0.1) |
| 1(P), OCRw | −0.0408 (−8.3) | −0.0195 (−4.0) | −0.0749 (−15.3) | −0.0020 (−0.4) | 0.0402 (9.6) | 0.0193 (4.7) | 0.0700 (19.9) | 0.0024 (0.6) |
| 1(P), IPWw | 0.0534 (10.9) | −0.0637 (−13.0) | −0.0876 (−17.9) | −0.0144 (−2.9) | −0.0004 (−1.0) | −0.0209 (−5.1) | 0.0169 (4.8) | −0.0138 (−3.3) |
| 2(P), OCR | −0.0004 (−0.1) | −0.0004 (−0.1) | 0.0002 (0.0) | 0.0001 (0.0) | −0.0002 (0.0) | −0.0003 (−0.1) | −0.0003 (−0.1) | −0.0002 (0.0) |
| 2(P), IPW | −0.0009 (−0.1) | 0.0011 (0.2) | −0.0002 (0.0) | 0.0007 (0.1) | −0.0003 (−0.1) | −0.0004 (−0.1) | −0.0005 (−0.1) | −0.0002 (0.0) |
| 3(P), OCR | −0.0000 (0.0) | −0.0008 (−0.4) | −0.0007 (−0.3) | −0.0010 (−0.5) | 0.0000 (0.0) | −0.0035 (−0.9) | −0.0025 (−0.7) | −0.0051 (−1.2) |
| 3(P), IPW | −0.0001 (−0.1) | −0.0011 (−0.5) | −0.0009 (−0.4) | 0.0004 (0.2) | −0.0030 (−1.1) | −0.2527 (−62.8) | −0.2195 (−58.1) | −0.2348 (−55.7) |
| 4(P), OCR | −0.0001 (0.0) | −0.0015 (−0.3) | 0.0001 (0.0) | −0.0003 (−0.1) | 0.0004 (0.1) | 0.0005 (0.1) | 0.0006 (0.1) | 0.0004 (0.1) |
| 4(P), IPW | 0.0000 (0.0) | −0.0014 (−0.3) | 0.0003 (0.1) | 0.0000 (0.0) | 0.0005 (0.1) | 0.0008 (0.1) | 0.0004 (0.1) | 0.0005 (0.1) |
| 5(P), OCR | −0.0056 (−0.8) | −0.0002 (−0.0) | −0.002 (−0.3) | −0.0001 (−0.0) | 0.2845 (93.1) | 0.0093 (1.5) | 0.0183 (3.1) | 0.0051 (0.8) |
| 5(P), IPW | −0.2143 (−29.9) | 0.0020 (0.27) | −0.0180 (−2.5) | 0.0237 (3.3) | −0.2380 (−77.9) | −0.0420 (−6.7) | −0.0996 (−16.8) | −0.0171 (−2.6) |
| 1(RR), OCR | 0.0109 (0.8) | 0.0035 (0.3) | 0.0121 (0.9) | 0.0054 (0.4) | 0.0042 (0.4) | 0.0016 (0.1) | 0.0029 (0.3) | 0.0040 (0.3) |
| 1(RR), IPW | 0.0189 (1.5) | 0.0781 (6.0) | 0.0879 (6.8) | 0.0243 (1.9) | 0.0089 (0.8) | 0.0340 (2.8) | 0.0262 (2.9) | 0.0161 (1.2) |
| 1(RR), OCRw | −0.2164 (−16.7) | −0.1169 (−9.0) | −0.3712 (−28.7) | −0.0086 (−0.7) | 0.2451 (22.8) | 0.1406 (11.7) | 0.4346 (47.7) | 0.0203 (1.5) |
| 1(RR), IPWw | 0.1018 (7.9) | −0.1651 (−12.7) | −0.1081 (−8.3) | −0.0209 (−1.6) | −0.0700 (−6.5) | −0.1869 (−15.6) | −0.1082 (−11.9) | −0.0672 (−5.1) |
| 2(RR), OCR | 0.0172 (0.8) | 0.0078 (0.5) | 0.0111 (0.5) | 0.0115 (0.6) | 0.0062 (0.6) | 0.0077 (0.5) | 0.0064 (0.5) | 0.0063 (0.5) |
| 2(RR), IPW | 0.0642 (3.0) | 0.0198 (1.1) | 0.0235 (1.1) | 0.0342 (1.8) | 0.0194 (1.8) | 0.0168 (1.2) | 0.0161 (1.4) | 0.0154 (1.2) |
| 3(RR), OCR | 0.0011 (0.2) | −0.0001 (−0.2) | −0.0006 (−0.1) | −0.0014 (−0.3) | 0.0027 (0.5) | −0.0057 (−0.7) | −0.0017 (−0.2) | −0.0089 (−1.1) |
| 3(RR), IPW | 0.0062 (1.3) | 0.0324 (6.8) | 0.0336 (7.0) | 0.0334 (7.0) | 0.0260 (4.9) | 0.4119 (48.2) | 0.4176 (50.9) | 0.3649 (45.2) |
| 4(RR), OCR | 0.0072 (0.5) | 0.0007 (0.1) | 0.0063 (0.5) | 0.0042 (0.4) | 0.0075 (0.5) | 0.0075 (0.5) | 0.0075 (0.5) | 0.0081 (0.5) |
| 4(RR), IPW | 0.0118 (0.8) | 0.0044 (0.3) | 0.0103 (0.8) | 0.0092 (0.8) | 0.0119 (0.7) | 0.0128 (0.8) | 0.0123 (0.8) | 0.0123 (0.8) |
| 5(RR), OCR | 0.0012 (0.1) | 0.0014 (0.2) | 0.0007 (0.1) | 0.0006 (0.1) | 0.2434 (44.2) | 0.0102 (1.2) | 0.0211 (2.7) | 0.0058 (0.7) |
| 5(RR), IPW | −0.0324 (−3.7) | 0.0064 (0.7) | 0.0018 (0.2) | 0.0104 (1.2) | 0.2627 (47.8) | 0.0295 (3.6) | 0.0503 (6.4) | 0.0174 (2.1) |

*Note.* In each cell, the first number represents the absolute bias, and the second number (in parentheses) represents the relative bias.

Abbreviations: IPW, inverse probability weighting approach with correctly specified propensity score model (except for setting 5); IPWw, inverse probability weighting approach with the propensity score model incorrectly specified (ie, by not including the essential covariate-covariate interaction term—setting 1); OCR, outcome regression approach with correctly specified outcome model; OCRw, outcome regression approach with the outcome model incorrectly specified (ie, by not including the essential treatment-covariate interaction term—setting 1); P, bias assessment on the probability scale; RR, bias assessment on the relative risk scale.

[a] 1 to 5: the setting.

**TABLE 3** Simulation results: summary estimates, between-trial variance, and $I^2$ statistics in the population-specific meta-analyses

| | OCR | | | IPW | | |
|---|---|---|---|---|---|---|
| **Setting[a]** | **Summary[b]** | **Between-Trial Variance[c]** | **$I^2$ Value[d]** | **Summary[b]** | **Between-Trial Variance[c]** | **$I^2$ Value[d]** |
| **1 (C)** | **0.14 (0.002)** | **0.019 (0.011, 0.030)** | **71 (57, 79)** | **0.14 (0.003)** | **0.018 (0.007, 0.033)** | **57 (34, 70)** |
| 1, $j = 1$ | 0.26 (0.002) | 0.0 (0.0, 0.002) | 0 (0, 24) | 0.25 (0.006) | 0.0 (0.0, 0.016) | 0 (0, 36) |
| 1, $j = 2$ | 0.07 (0.002) | 0.0 (0.0, 0.002) | 0 (0, 21) | 0.07 (0.006) | 0.0 (0.0, 0.011) | 0 (0, 30) |
| 1, $j = 3$ | 0.18 (0.002) | 0.0 (0.0, 0.003) | 0 (0, 22) | 0.17 (0.006) | 0.0 (0.0, 0.012) | 0 (0, 32) |
| 1, $j = 4$ | −0.09 (0.003) | 0.0 (0.0, 0.003) | 0 (0, 22) | −0.09 (0.006) | 0.0 (0.0, 0.012) | 0 (0, 31) |
| 1, $j = 5$ | 0.28 (0.002) | 0.0 (0.0, 0.003) | 0 (0, 23) | 0.27 (0.006) | 0.0 (0.0, 0.010) | 0 (0, 27) |
| 1.1, $j = 1$ | 0.13 (0.001) | 0.017 (0.009, 0.026) | 72 (59, 79) | 0.22 (0.004) | 0.002 (0.0, 0.015) | 11 (0, 44) |
| 1.1, $j = 2$ | 0.14 (0.002) | 0.020 (0.011, 0.029) | 72 (59, 79) | 0.03 (0.004) | 0.0 (0.0, 0.009) | 0 (0, 32) |
| 1.1, $j = 3$ | 0.15 (0.002) | 0.022 (0.012, 0.033) | 72 (59, 79) | 0.14 (0.004) | 0.003 (0.0, 0.017) | 13 (0, 46) |
| 1.1, $j = 4$ | 0.15 (0.002) | 0.022 (0.013, 0.034) | 72 (59, 79) | −0.13 (0.005) | 0.0 (0.0, 0.012) | 0 (0, 36) |
| 1.1, $j = 5$ | 0.15 (0.002) | 0.022 (0.012, 0.033) | 72 (59, 79) | 0.24 (0.004) | 0.0 (0.0, 0.009) | 0 (0, 31) |
| **2 (C)** | **0.45 (0.001)** | **0.0 (0.0, 0.003)** | **0 (0, 30)** | **0.45 (0.002)** | **0.0 (0.0, 0.004)** | **0 (0, 27)** |
| 2, $j = 1$ | 0.61 (0.002) | 0.013 (0.006, 0.022) | 62 (44, 73) | 0.60 (0.005) | 0.011 (0.0, 0.030) | 38 (0, 61) |
| 2, $j = 2$ | 0.30 (0.002) | 0.018 (0.010, 0.027) | 71 (59, 79) | 0.31 (0.004) | 0.014 (0.001, 0.031) | 42 (7, 62) |
| 2, $j = 3$ | 0.54 (0.002) | 0.013 (0.007, 0.021) | 65 (49, 75) | 0.53 (0.004) | 0.011 (0.0, 0.026) | 39 (0, 60) |
| 2, $j = 4$ | 0.35 (0.002) | 0.017 (0.010, 0.026) | 71 (58, 79) | 0.35 (0.004) | 0.014 (0.002, 0.030) | 46 (12, 64) |
| 2, $j = 5$ | 0.40 (0.002) | 0.015 (0.008, 0.022) | 70 (56, 78) | 0.40 (0.004) | 0.011 (0.0, 0.026) | 40 (1, 61) |
| **3 (C)** | **−0.35 (0.002)** | **0.040 (0.030, 0.050)** | **85 (81, 88)** | **−0.35 (0.003)** | **0.042 (0.026, 0.061)** | **76 (66, 81)** |
| 3, $j = 1$ | −0.74 (0.005) | 0.0 (0.0, 0.006) | 0 (0, 22) | −0.76 (0.019) | 0.0 (0.0, 0.044) | 0 (0, 35) |
| 3, $j = 2$ | −0.59 (0.004) | 0.0 (0.0, 0.002) | 0 (0, 15) | −0.60 (0.013) | 0.0 (0.0, 0.025) | 0 (0, 31) |
| 3, $j = 3$ | −0.16 (0.003) | 0.0 (0.0, 0.0) | 0 (0, 0) | −0.20 (0.010) | 0.011 (0.0, 0.034) | 33 (0, 60) |
| 3, $j = 4$ | −0.19 (0.003) | 0.0 (0.0, 0.0) | 0 (0, 0) | −0.23 (0.009) | 0.010 (0.0, 0.030) | 32 (0, 59) |
| 3, $j = 5$ | −0.21 (0.003) | 0.0 (0.0, 0.0) | 0 (0, 0) | −0.26 (0.010) | 0.011 (0.0, 0.034) | 28 (0, 56) |
| **4 (C)** | **0.32 (0.001)** | **0.015 (0.008, 0.023)** | **68 (52, 77)** | **0.31 (0.003)** | **0.014 (0.004, 0.026)** | **52 (23, 67)** |
| 4, $j = 1$ | 0.32 (0.002) | 0.015 (0.007, 0.023) | 68 (51, 76) | 0.31 (0.003) | 0.013 (0.004, 0.025) | 51 (22, 66) |
| 4, $j = 2$ | 0.32 (0.002) | 0.015 (0.007, 0.023) | 68 (51, 76) | 0.31 (0.003) | 0.013 (0.004, 0.026) | 51 (23, 66) |
| 4, $j = 3$ | 0.32 (0.002) | 0.015 (0.007, 0.023) | 68 (51, 76) | 0.31 (0.003) | 0.013 (0.004, 0.025) | 51 (22, 66) |
| 4, $j = 4$ | 0.32 (0.002) | 0.015 (0.007, 0.023) | 68 (51, 76) | 0.31 (0.003) | 0.013 (0.004, 0.025) | 51 (23, 66) |
| 4, $j = 5$ | 0.32 (0.002) | 0.015 (0.007, 0.023) | 68 (51, 76) | 0.31 (0.003) | 0.013 (0.004, 0.026) | 51 (22, 66) |
| **5 (C)** | **−0.24 (0.001)** | **0.016 (0.012, 0.020)** | **87 (84, 89)** | **−0.24 (0.002)** | **0.024 (0.015, 0.035)** | **74 (64, 81)** |
| 5, $j = 1$ | −0.12 (0.001) | 0.0 (0.0, 0.0) | 0 (0, 25) | −0.12 (0.002) | 0.0 (0.0, 0.004) | 0 (0, 27) |
| 5, $j = 2$ | −0.31 (0.009) | 0.051 (0.026, 0.080) | 89 (80, 91) | −0.36 (0.005) | 0.024 (0.007, 0.045) | 52 (25, 67) |
| 5, $j = 3$ | −0.18 (0.001) | 0.0 (0.0, 0.0) | 0 (0, 26) | −0.18 (0.002) | 0.0 (0.0, 0.004) | 0 (0, 27) |
| 5, $j = 4$ | −0.18 (0.001) | 0.0 (0.0, 0.0) | 0 (0, 26) | −0.18 (0.002) | 0.0 (0.0, 0.004) | 0 (0, 27) |
| 5, $j = 5$ | −0.17 (0.001) | 0.0 (0.0, 0.0) | 0 (0, 25) | −0.17 (0.002) | 0.0 (0.0, 0.004) | 0 (0, 27) |

Abbreviations: IPW, inverse probability weighting approach; IQR, interquartile range; OCR, outcome regression approach.

[a]1(C) to 5(C), in bold: results of the standard two-step meta-analysis in each setting (from setting 1 to setting 5); 1, j to 5, j: results of the population-j-specific meta-analysis (j=1,...,5) in each setting (from setting 1 to 5), when the models involved in the OCR and IPW estimators are correctly specified (except for the IPW estimators in setting 5), 1.1,j: results of the population-j-specific meta-analysis in setting 1 when the models involved in the OCR and IPW estimators are incorrectly specified.

[b]Summary estimate (variance).

[c]Median (IQR).

[d]Median (IQR).

**TABLE 4** Simulation results: heterogeneity tests (the percentage of simulations showing statistically significance at a type I error risk of 5%)

| | | Setting 1 | | | | Setting 2 | | Setting 3 | | Setting 4 | Setting 5 | | |
|---|---|---|---|---|---|---|---|---|---|---|---|---|---|
| | | OCR | IPW | OCRw | IPWw | OCR | IPW | OCR | IPW | OCR | IPW | OCR | IPW |
| New approach: beyond case-mix heterogeneity tests[a] | $j=1$ | 4.4 | 8.1 | 76.5 | 12.5 | 60.4 | 29.5 | 5.2 | 9.3 | 70.1 | 41.2 | 4.3 | 5.9 |
| | $j=2$ | 4.6 | 6.3 | 76.4 | 7.4 | 78.8 | 32.2 | 5.1 | 7.7 | 70.2 | 41.2 | 87.2 | 45.1 |
| | $j=3$ | 4.5 | 6.3 | 76.4 | 14.0 | 64.9 | 29.2 | 4.1 | 30.0 | 70.1 | 41.1 | 5.2 | 5.3 |
| | $j=4$ | 4.8 | 6.2 | 76.1 | 9.1 | 77.1 | 35.3 | 4.1 | 28.5 | 70.1 | 41.1 | 5.1 | 5.5 |
| | $j=5$ | 4.3 | 5.7 | 76.6 | 6.5 | 75.4 | 30.2 | 4.3 | 25.4 | 70.1 | 41.2 | 5.1 | 5.4 |
| New approach: case-mix heterogeneity tests[b] | $k=1$ | 100 | 87.0 | 33.3 | 88.7 | 99.9 | 77.9 | 97.7 | 36.1 | 1.8 | 0.3 | 22.4 | 2.5 |
| | $k=2$ | 99.8 | 81.7 | 0.1 | 92.7 | 99.4 | 56.3 | 100 | 29.2 | 1.8 | 0.2 | 97.0 | 31.6 |
| | $k=3$ | 100 | 98.2 | 4.8 | 97.3 | 100 | 96.7 | 99.9 | 28.8 | 1.7 | 0.4 | 47.6 | 4.5 |
| | $k=4$ | 100 | 94.0 | 2.8 | 91.6 | 99.9 | 82.8 | 99.6 | 28.9 | 1.9 | 0.5 | 48.0 | 3.3 |
| | $k=5$ | 99.9 | 74.1 | 13.2 | 78.4 | 99.8 | 72.0 | 99.9 | 30.4 | 1.9 | 0.5 | 44.7 | 5.2 |
| Conventional heterogeneity test[c] | $j=k$ | 74.3 | 48.9 | 74.2 | 48.9 | 6.2 | 5.9 | 99.0 | 85.8 | 67.9 | 40.6 | 99.7 | 82.7 |

Abbreviations: IPW, inverse probability weighting approach with correctly specified propensity score model (except for setting 5); IPWw, inverse probability weighting approach with the propensity score model incorrectly specified (ie, by not including the essential covariate-covariate interaction term—setting 1); OCR, outcome regression approach with correctly specified outcome model; OCRw, outcome regression approach with the outcome model incorrectly specified (ie, by not including the essential treatment-covariate interaction term—setting 1).

[a]Tests comparing $\log RR(j,k)$ with the same value of $j$.

[b]Tests comparing $\log RR(j,k)$ with the same value of $j$.

[c]Tests comparing $\log RR(j,j)$ ($j,k = 1,\ldots,5$).

presence of heterogeneity but is unable to indicate that such heterogeneity is due to the differential case mix across studies (Tables 3 and 4).

In setting 1, the IPW-based tests assessing beyond case-mix heterogeneity are slightly inflated in terms of type I error (Table 4), which results from the minor bias in the IPW-based estimates that have been discussed above. As the bias is small, the between-trial variance and $I^2$ statistics of the population-specific meta-analyses are still properly estimated. In contrast, the heterogeneity assessment becomes inaccurate when the outcome model for the OCR-based approach or the PS model for the IPW-based approach is severely misspecified.

In setting 3, results of the meta-analysis and subsequent heterogeneity assessment specific for trials 3 to 5 are biased when using the IPW-based estimator. Similar to setting 1, the beyond case-mix heterogeneity tests using IPW-based estimates are slightly inflated in type I error when the population of interest is of trials 1 and 2 (Table 4). The between-trial variance and $I^2$ statistics in the meta-analyses specific for populations 1 and 2 still correctly specify that no beyond case-mix heterogeneity is present (Table 3).

In setting 4, the tests assessing case mix heterogeneity are likely too conservative. The Wald statistics in these tests are shrunken toward 0, which makes the type I error lower than the conventional level of 5% (Table 4). The reason is that the estimates $\log \hat{RR}(j,k)$ with the same $k$ are strongly correlated under the null hypothesis of no case-mix heterogeneity, which makes the matrix $\left[\boldsymbol{M}\left(\hat{\sum}\right)\boldsymbol{M}^t\right]$ in the Wald statistic have extremely small entries. In such a situation, a slight bias in estimating the covariance matrix $\sum$ can result in a considerable bias on the Wald statistic. More sophisticated methods for variance estimation may then be indicated. This may however not be a major practical concern as we did not observe it in any other settings, in which the distributions of the covariates $L_1$ to $L_5$ across the five populations are truly heterogeneous.

In setting 5, only the results of the meta-analysis and subsequent heterogeneity assessment specific for trial 2 give valid results when using the OCR-based estimator. In contrast, results are invalid when using the IPW-based estimator due to the bias discussed above.

In summary, both the OCR-based and IPW-based estimators are effective for case-mix standardization across different populations. They enable a more accurate and refined heterogeneity assessment as compared with a standard meta-analysis and deliver a pooled estimate that expresses the effect for a well-defined patient population. The OCR-based estimator is optimal if the outcome model in each population is correctly specified. However, model misspecification is likely and difficult to diagnose when the different studies have very different case mix. Therefore, when using this estimator, it is best based on a flexible model that incorporates many possible treatment-covariate and covariate-covariate interactions, regardless of whether there is significant evidence for them. To lessen the risk of overfitting, the IPW approach seems more promising. First, it does not require modeling treatment-covariate interactions. Second, when the different trials in the meta-analysis are at least similar on the PICO basis, then positivity violations and thus extreme weights are less likely. Future frameworks should focus on improving the performance of this IPW approach.

## 4 | META-ANALYSIS OF THE EFFECT OF VITAMIN D SUPPLEMENTATION ON ACUTE RESPIRATORY TRACT INFECTION

We apply the proposed approach to a recently published IPD meta-analysis assessing the overall effect of vitamin D supplementation on the risk of experiencing at least one acute respiratory tract infection.[26] Data for six eligible trials that include information on six baseline covariates are available for analysis (Supporting Information S4). For this illustration, we only consider the covariates that were collected across all trials. These are gender, age, body mass index (BMI), influenza vaccination status, and vitamin D concentration at baseline. All six trials adopt a randomization ratio of 1:1. One trial is a cluster randomized trial, and one other has a relatively small sample size (ie, 34 participants). We exclude the small trial and, for this illustration, ignore the potential clustering effect in the cluster randomized trial. The target population of one trial was moreover found to be very different from the others; ie, it only includes male participants with 18 to 21 years of age (Supporting Information S4). To prevent potential violations of positivity, this trial is excluded from the meta-analysis, leaving four trials.

We apply the new IPW approach to meta-analyze the dataset on the log odds ratio scale. The main terms of all baseline covariates are included in the multinomial PS model. To decide on the inclusion of two-way interactions, we run two independent backward elimination processes: one for the logistic outcome model and the other for the multinomial PS model. Any interaction term that is included in one of the two final sets is considered for adjustment in the meta-analysis. This approach leads to the inclusion of five interaction terms, namely, between sex and BMI, between sex and flu vaccination status, between age and gender, between age and influenza vaccination status, and between age and BMI.

For each IPW estimate, the weights are truncated by resetting the value of weights greater than the 95th percentile to the value of the 95th percentile. The presence of large weights after truncation (ie, higher than 200) indicates a potential violation of the positivity assumption. If this is the case, we keep the corresponding IPW estimate in the meta-analysis to see its impact on the final summary.

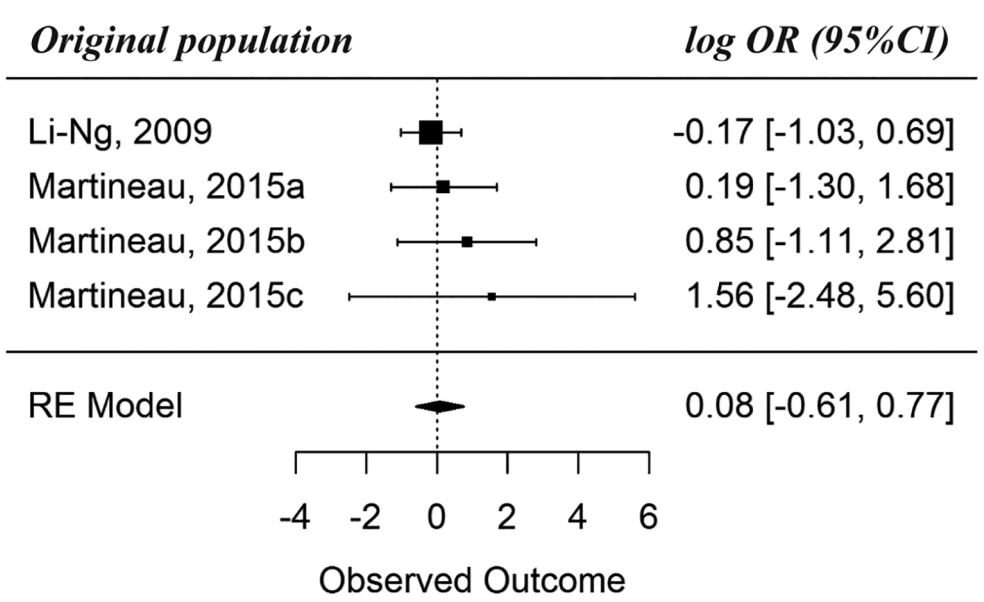

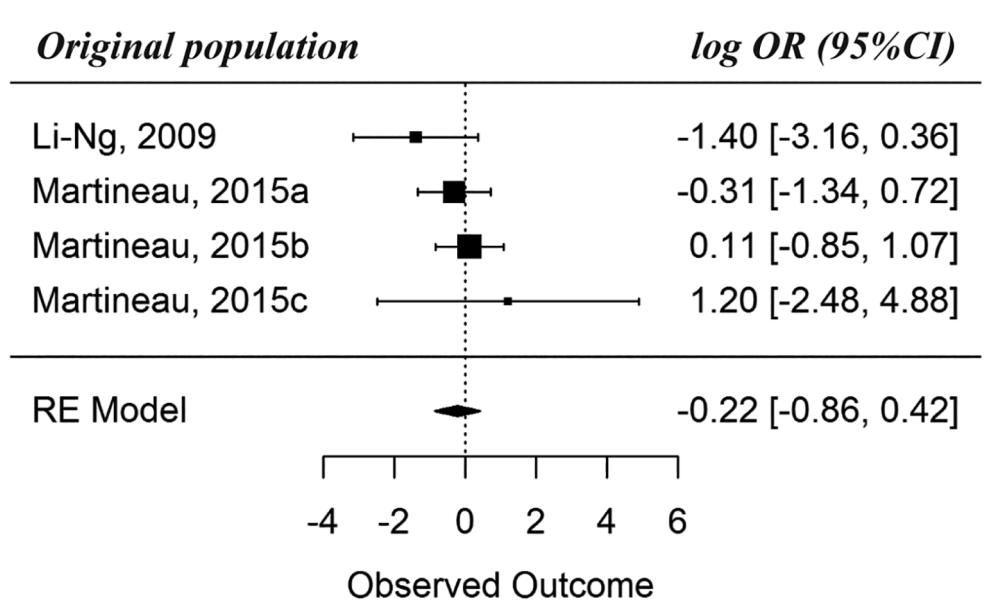

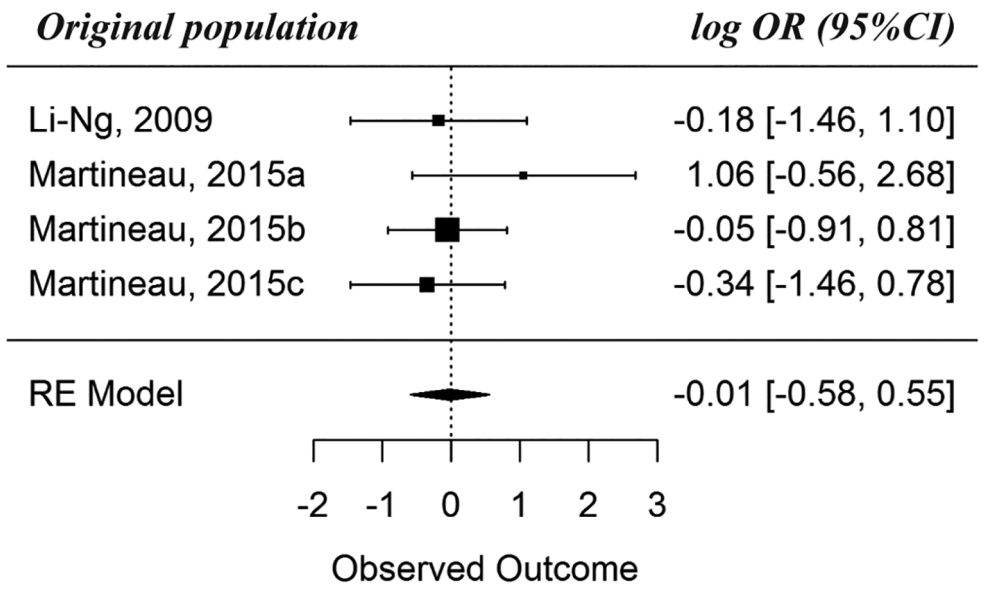

**FIGURE 2** Data analysis: the population-specific meta-analyses

Results of the population-specific meta-analyses are given in Figure 2 and Supporting Information S5. The weight distributions after truncation are shown in Supporting Information S6. As can be seen from these weights, there is a clear violation of positivity when transporting the result from trial 4 (Martineau, 2015c) to trial 2 (Martineau, 2015a). This gets translated into a large standard error for the corresponding IPW estimate for $OR(2,4)$ (Supporting Information S6), whose impact on the final result is therefore dampened.

The population-specific summary effects are not statistically significant across different trial populations. However, vitamin D tends to be less effective than placebo in population 2, and the two treatments appear equally effective in the other populations, although these findings are not statistically significant (possibly due to the lack of power). Besides, there is no statistically significant evidence of heterogeneity, neither due to case mix nor due to beyond case mix (Supporting Information S5). Finally, since the IPD were only obtained for six trials, the findings reported here might be subject to selection bias. We thus illustrate the proposed approach but do not aim to make clinical inference.

## 5 | DISCUSSION

Assessing the impact of case-mix variation across the eligible studies is an important task in every meta-analysis. Case-mix heterogeneity, when it exists, can be quite a nuisance as it can make the result from different trials difficult to pool. In this paper, we propose a novel framework that standardizes evidences across different trials to one well-defined population before summarizing them. Simulation results demonstrate the adequacy of the new approach and indicate that such an approach allows for a more informative heterogeneity assessment. Dismantling case-mix heterogeneity from the total heterogeneity is especially meaningful since case-mix heterogeneity and beyond case-mix heterogeneity may sometimes compensate each other, thereby resulting in approximately equal marginal effect estimates (eg, see setting 2 of the simulation study).

Our proposal is readily extended to meta-analyses of observational studies. In the OCR approach, this merely requires that the OCR model additionally includes confounders of the treatment-outcome association. It is just slightly more involved in the IPW approach, where this would require additional weighting by the reciprocal of the probability of the observed treatment, given confounders. The resulting procedure for observational studies is arguably of even greater importance. Here, different studies typically adjust for different covariate sets, which may result in excess heterogeneity. Indeed, even if all studies evaluated the same study population and controlled for a sufficient set of confounders, typical effect measures (such as odds ratios and hazard ratios) would differ systematically between studies when some adjust for additional covariates and others do not. This is the result of noncollapsibility of these effect measures.[11,17,27,28] It can make the treatment effects from different observational studies difficult to pool. The proposed procedure overcomes this by standardizing the results from all studies to the same population.

We did not discuss a number of important issues, such as the problem of covariates being systematically missing in some trials or how to take into account the trials with limited sample size or with special study designs (eg, clustered or noninferiority trials). A relatively large sample size was also chosen in the simulation study, as the primary objective was to investigate the validity of the new meta-analysis and heterogeneity assessment approaches under reasonably good conditions in terms of power. The new approaches, therefore, should be further evaluated in a wider range of settings and of various sample sizes. Further, as individual patient data can be difficult to obtain in practice, it is important that the proposed approaches can be extended to aggregated data, so as to make it more data-friendly and more widely applicable.

Finally, a drawback of the proposed approaches is that they require different random-effects meta-analyses, each targeted to the population of a different trial. This can easily be avoided, however, by instead standardizing the results to the population of only one of the trials $j$. From the viewpoint of generalizability, this is ideally the trial with the most heterogeneous case mix. To avoid positivity violations, this is ideally the trial with the tightest case mix. Alternatively, one may standardize the results to the population observed in an external reference electronic health registry. As noted by a referee, trialists may then consider mutual standardization in the original trial reports. In particular, each trial might then use inverse probability weighting to produce an effect measure estimate standardized to the case-mix distribution in that reference register. Then, meta-analysts could base a standard meta-analysis on these mutually standardized estimates, which would have the advantage of describing the effect for the same population. This would overcome the need for an IPD meta-analysis. As an added advantage, this may often lead to a reduction in between-trial heterogeneity. In practice, such an approach is also useful for supporting the decision-making process. For instance, public health authorities in a given population might consider standardizing results of the different trials conducted elsewhere over the realistic case mix encountered in their population. Results of the meta-analysis after this standardization will reflect more honestly the treatment effectiveness for such population structure, which is informative to decide whether or not the new intervention should be recommended in the interested population. While such a strategy contributes to increase the generalizability of the findings by directly addressing the issues of case mix, it does not take into account other types of non-generalizability, such as the one arising from differences between the real and anticipated trial interventions. To address this, the estimates $RR(j,k)$ from trials $k$ that come closest to being "pragmatic" are likely the ones that should receive most emphasis in the final meta-analysis. We will investigate this in future work.

To summarize, we developed a novel meta-analysis approach for randomized clinical trials, which uses individual patient data from all trials to infer the treatment effect for the patient population in a given trial, based on either OCR or IPW. We investigated the new approach via numerically simulated data and showed that the new approach can lead to insightful heterogeneity assessment in practice. Via reanalyzing the real dataset of a published IPD meta-analysis, we also showed that the new approach is applicable in practice.

## ORCID

*Tat-Thang Vo* https://orcid.org/0000-0001-6485-9947

## SUPPORTING INFORMATION

Additional supporting information may be found online in the Supporting Information section at the end of this article.